\documentclass[twocolumn,prl,showpacs,floatfix]{revtex4}
\usepackage{graphicx}
\usepackage{dcolumn}
\usepackage{amsmath}
\usepackage{latexsym}
\begin{document}

\title{Driven Disordered Periodic Media with an Underlying Structural
Phase Transition}

\author{Ankush Sengupta and Surajit Sengupta}
\affiliation{
Satyendra Nath Bose National Centre for Basic Sciences, Block-JD, 
Sector-III, Salt Lake, Kolkata 700 098, India
}
\author{Gautam I. Menon}
\affiliation{
The Institute of Mathematical Sciences, CIT Campus, 
Taramani, Chennai 600 113, India
}
\date{\today}
\begin{abstract}

We investigate the driven states of a two-dimensional
crystal whose ground state can be tuned through a
square-triangular transition. The depinning of such a
system from a quenched random background potential occurs
via a complex sequence of dynamical states, which include
plastic flow states, hexatics, dynamically stabilized
triangle and square phases and intermediate regimes of
phase coexistence. These results are relevant to transport
experiments in the mixed phase of several superconductors
which exhibit such structural transitions as well as to
driven colloidal systems whose interactions can be tuned
via surface modifications.

\end{abstract} 
\maketitle

The description of phases and phase transitions
in driven steady states is a central theme in
the statistical mechanics of non-equilibrium
systems\cite{special}.  A variety of such states
are obtained in the depinning and flow of randomly
pinned, {\em periodic} media, such as charge-density
wave systems and Abrikosov flux-line lattices in the
mixed state of type-II superconductors\cite{gia}.
Many such superconductors exhibit structural
transitions in the mixed state, typically between
flux-line lattices with triangular and rectangular
symmetry \cite{super}. Colloidal systems such as
PMMA spheres coated with a low-molecular-weight
polymer undergo a remarkable variety of solid-solid
transformations in an external field\cite{anand}. While
applying a sufficiently large current depins the
flux lines from the quenched random disorder present
in all real materials, the possibility of 
driving colloidal particles in two dimensions across a
disordered substrate has also been raised\cite{coll-rand}.
What links these diverse systems is the generic problem
of understanding the competition between an underlying
structural phase transition in a pure periodic system
as modified by disorder, and the non-equilibrium
effects of an external drive. This Letter proposes and
studies a simple model which describes this physics.

\begin{figure}[t]
\begin{center}
\includegraphics[width=8.0cm]{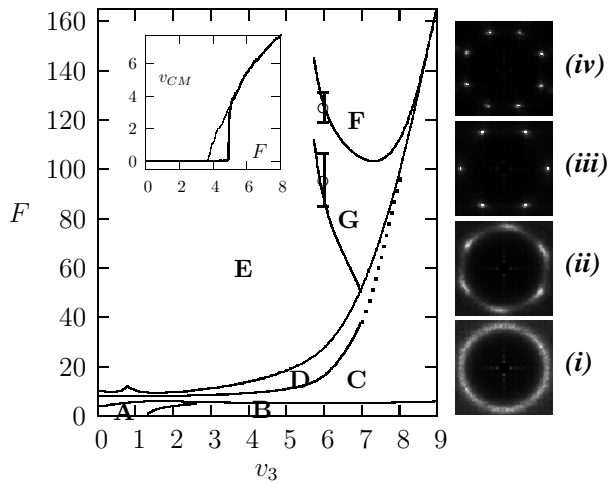}
\end{center}
\caption{
The dynamical phase diagram in the $v_3 - F$
plane.  The phases are: A --
pinned triangle, B -- pinned square, C -- plastic
flow/isotropic liquid, D -- moving hexatic,
E -- moving triangle, F -- moving square and G  --
dynamical square-triangle coexistence.  Two points
with error bars show the boundary of G for $v_3 = 6$,
obtained by averaging over 24 disorder realizations,
as an example.  The boundaries for all other
transitions are considerably sharper.  The inset
shows the centre of mass velocity $v_{cm}$ as
a function of the driving force $F$ as the force is
increased (bold) and then decreased across the depinning
transition.  The right panel shows
$S({\bf q})$ for the plastic flow state
(i), hexatic glass (ii), triangular solid (iii) and 
square solid phases (iv) at $v_3 = 6.0$ and $F = 10, 20,
60$ and $140$ respectively. To obtain $S({\bf q})$, $50$
independent configurations
were used. The structure in (iv) reflects the presence of
two mutually misoriented square crystallites.
} 
\label{pdia}
\end{figure}

Our model system is two-dimensional and
consists of particles with two and three-body
interactions\cite{2dmod}. The three-body interaction,
parametrized through a single parameter $v_3$, tunes
the system across a square-triangular phase transition.
Our central result, the sequence of steady states
obtained as a function of increasing force $F$ for
various values of $v_3$, is summarized in the dynamical
``phase'' diagram of Fig.~\ref{pdia}.  We obtain
a variety of phases: pinned states which may have
dominantly triangular or square correlations, a moving
liquid/glass phase, a moving hexatic glassy phase,
flowing triangular and square states ordered over
the size of our simulation cell and a dynamic
coexistence regime between these ordered phases. We
discuss our characterization of these states and
the applicability of simple dynamical criteria for
non-equilibrium phase transitions between them.

{\em The model:\,}      
Particles interact in two dimensions through
the interaction potential $1/2\sum_{i\ne j}V_2(r_{ij})+
1/6\sum_{i \ne j \ne k} V_3(r_{i},r_{j},r_{k})$, where ${\bf r}_i$ is the
position vector of particle $i$, $r_{ij} \equiv |{\bf r}_{ij}| \equiv 
|{\bf r}_j-{\bf r}_i|$,
$V_2(r_{ij})=v_{2}(\frac {\sigma_0}{r_{ij}})^{12}$
and $V_3(r_{i},r_{j},r_{k})=v_{3}[f_{ij}sin^{2}(4\theta_{ijk})f_{jk}+{\rm permutations}]$\cite{2dmod}.
The function $f_{ij} \equiv f(r_{ij}) = (r_{ij}- r_{0})^2$ for
$r_{ij}<1.8 \sigma_0$ and $0$ otherwise and $\theta_{ijk}$ is the angle
between ${\bf r}_{ji}$ and ${\bf r}_{jk}$.
The two-body (three-body) interaction favours a triangular (square)
ground state.
Energy and length scales are set using $v_2 = 1$ and   
$\sigma_0 = 1$.
Particles also interact with a quenched random
background modeled as a Gaussian random 
potential\cite{Chud} 
V$_d({\bf
r})$ with zero
mean and exponentially decaying (short-range) correlations, 
The disorder variance is
set to $v_d^2 = 1$ and its spatial correlation length is
$\xi=0.12$. 
The system evolves through standard Langevin dynamics;
$\dot{{\bf r}_i} = {\bf v}_i$ and $\dot{{\bf v}_i}
= {{\bf f}_i}^{\rm int} - \alpha {\bf v}_i + {\bf F} + {\bf
\eta}_i(t)$. Here ${\bf v}_i$ is the velocity, ${{\bf
f}_i}^{\rm int}$ the total interaction force, and ${\bf
\eta}_i(t)$ the random force acting on particle $i$. A
constant force ${\bf F} = \{F_x,0\}$ 
drives the system and the zero-mean thermal noise ${\bf \eta}_i(t)$
is specified by $<{\bf \eta}_i(t){\bf
}{\bf \eta}_j(t')> = 2T\delta_{ij}\delta(t-t')$ with 
$T=0.1$, well below the equilibrium melting temperature of
the system. The unit of time $\tau = 
\alpha \sigma_0^2/v_2$, with $\alpha = 1$ the viscosity. 

{\em Simulation details:\,} Our system consists
of $1600$ particles in a square box 
at number density $\rho
= 1.1$. At $T = 0.1$, the pure system remains triangular
upto $v_3 = 1.5$. For larger $v_3$,
a square phase is obtained. 
Larkin length estimates\cite{gia} yield
$L_a/a  \sim 100$, 
with $a = 1/{\rho}^{1/2}$ the lattice parameter,
somewhat larger than our system size. 

Equilibration in the disorder potential is
achieved through a simulated annealing
procedure
in which the disorder potential is varied by
increasing its strength from zero to unity in small steps.
The system is stabilized for a minimum of $2\times10^5$
Monte Carlo steps at each disorder strength. Such
annealed configurations are our
initial inputs to the Langevin simulations. We evolve
the system using a time step of $10^{-4}\tau$.
The external force $F$ is ramped up from a
starting value of 0, with the
system maintained at upto $10^8$ steps at 
each $F$.  
\begin{figure}[t]
\begin{center} \includegraphics[width=8.0cm]{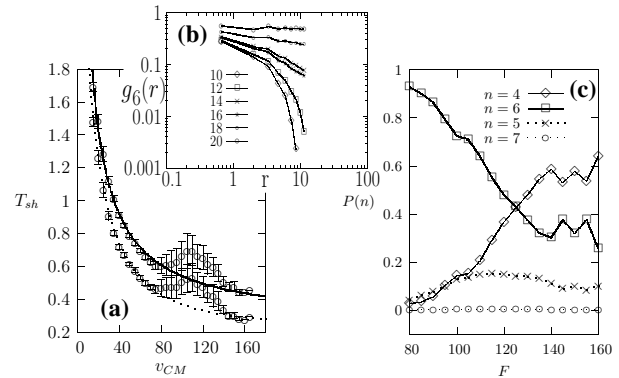}
\end{center} 
\caption{(a) The shaking temperatures for a
system with $v_3 = 6$ as a function of $F$ in
the drive ($x$) and transverse ($y$) directions.
These are
obtained by averaging over $100$ independent configurations
as well as over $25$ separate disorder
realizations.  For $F$ in the coexistence
region there is a significant enhancement of the shaking
temperature in excess of the prediction of 
KV (Ref.\cite{vinok}). The lines represent fits to the KV form.
(b) The hexatic correlation function $g_6(r)$ at varying force
values at $v_3=6$, averaged over 50 independent configurations,
(c) $P(n)$ for $n = 4,5,6,7$ (see text)
{\it vs} $F$ averaged as in (a). 
}
\label{shakord}
\end{figure}
\begin{figure}[t]
\begin{center} \includegraphics[width=4.0cm]{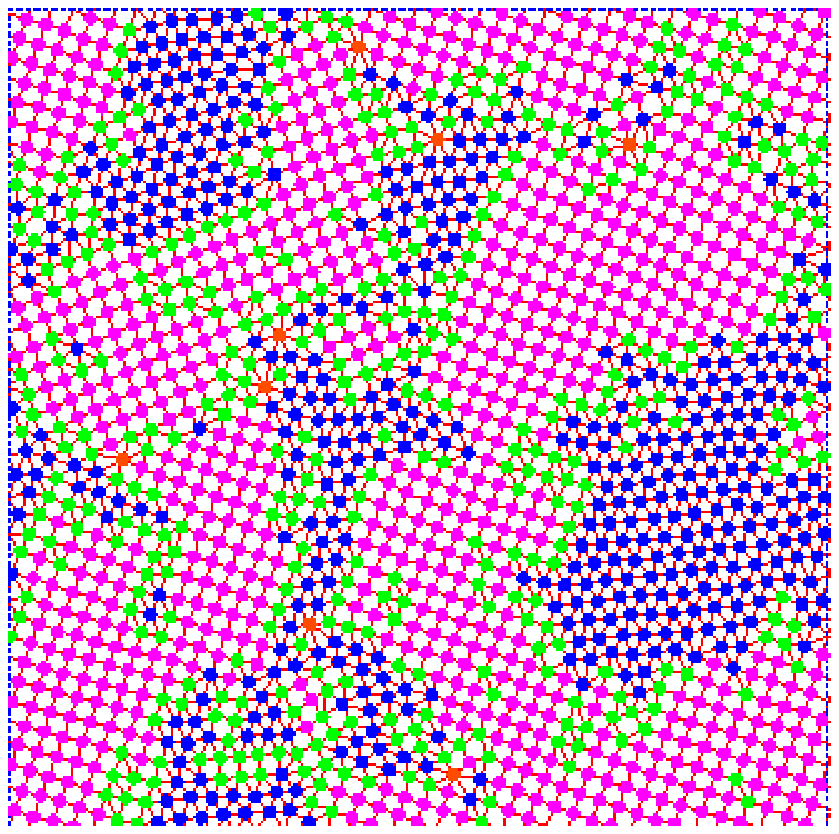}
\end{center} 
\caption{(color-online) A single particle configuration
showing square-triangle coexistence at $F=107$ together with the computed 
Delaunay mesh. The particles are 
colored according to the number of neighbors n = 4 (magenta), 5 (green),
6 (blue) and 7 (orange). Particles with coordination 5 
are present mainly in the interfacial region while those with 7 are
associated with isolated dislocations.}
\label{config}
\end{figure}

{\em Observables:\,} We monitor structural observables,
such as the static
structure factor $S({\bf q}) = \sum_{ij} \exp(-i{\bf q \cdot r}_{ij})$.
Delaunay triangulations yield
the probability distributions $P(n)$ 
of $n=4,5,6$ and $7$ coordinated 
particles ($\sum_n P(n) = 1$)\cite{delaunay}. We define 
order parameters
$\psi = (P(4)-P(6))/(P(4) + P(6))$ to distinguish between square 
and triangular phases, $\psi_{\Delta} = (P(6)-P(5)-P(7))/(P(6)+P(5)+P(7))$ to 
distinguish between liquid (disordered) and triangular crystals and 
$\psi_{\Box} = (P(4)-P(5)-P(7))/(P(4)+P(5)+P(7))$ to distinguish between 
liquid and square crystals. In addition, we compute the  
hexatic order parameter $\psi_6 = \sum_{ij} \exp(-i6\theta_{ij})$
and its correlations, where 
$\theta$ is the bond angle measured with respect to an 
arbitrary external axis.
The dynamical variables we study include the center
of mass velocity $v_{cm}$, the particle flux and its
statistics and the Koshelev-Vinokur (KV) ``shaking
temperature'' $T^\alpha_a$\cite{vinok} appropriate
to the drive and transverse directions and obtained from $T^\alpha_s =
\langle \left [ v^{\alpha} - v^{\alpha}_{cm}\right ]^2 \rangle /2$, $\alpha = x,y$.

\begin{figure}[t]
\begin{center}
\includegraphics[width=8.0cm]{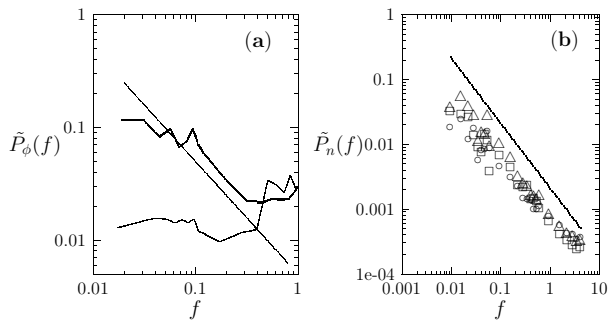}
\end{center}
\caption{(a) The power spectrum $\tilde{P}_\phi(f)$
of current fluctuations in the moving triangle phase (thin line $F=55$) 
and in the coexistence region (bold line $F=100$), logarithmically
binned and plotted for a range of frequencies below the washboard 
frequency.  Current fluctuations in the coexistence region
are {\em enhanced}, also showing a $1/f$ decay at intermediate frequencies. 
(b) $\tilde{P}_n(f)$  for 
the fluctuations of the numbers of $4 (\Box),5(\circ)$ and $6(\Delta)$ 
coordinated particles in the coexistence region.
The number of $7$ coordinated particles (not shown) is vanishingly small. 
The straight line in both figures
represents the $1/f$ law. }
\label{autocor}
\end{figure}

For small $F$ the solid is pinned.
A disorder broadened version of the equilibrium
triangle (A) to square (B) transition results as $v_3$
is varied across the $T=0$ transition value at small F;
here and below alphabets in brackets refer to the
states labelled in Fig.~\ref{pdia}.
The triangular (A) phase is favoured at non-zero $F$.
Upon further increasing $F$, the system undergoes a {\em
discontinuous} depinning transition which exhibits
prominent hysteresis behaviour (Fig.~\ref{pdia}, inset).
Such a depinned state is inhomogeneous and undergoes
plastic flow\cite{jensen,moon,faleski,vort-dyn,fangohr,chandran} 
consistent with earlier numerical work.
For larger $F$ the velocity approaches the
asymptotic behaviour $v_{CM} = F$.

The structure factor $S(q)$ of the plastically
moving phase (C) obtained just
above the depinning transition consists of
liquid-like isotropic rings
(Fig.~\ref{pdia}(i)). 
Upon increasing $F$, the circular 
ring in $S(q)$ concentrates into
six smeared peaks which we associate with a hexatic 
glass (D)\cite{dpt-hexatic}. Fig.~\ref{shakord}(b) shows
the evolution of the hexatic correlation function
$g_6(r) = \langle \psi_6(0)\psi_6(r)\rangle$, as $F$ is
varied across $(C) \to (D) 
\to (E)$. Note the sharp exponential
decay of hexatic correlations in (C), the
power-law or quasi-long-range order (QLRO)
decay in (D) and the saturation (LRO) of
this correlation function in regime (E).
The plastic flow regime (C) expands at larger
$v_3$\cite{largev3}.  On further increasing $F$, the
hexatic glass {\em recrystallizes} into a structure
which depends on the value of $v_3$: for low $v_3$
the final crystal is triangular (E) whereas for
large $v_3$ it is square (F). For intermediate $v_3$
the system first freezes into a triangular structure
but subsequently transforms into the square via an intervening
``coexistence'' regime (G) best
described as a mosaic of dynamically fluctuating square
and triangular regions.

The phase diagram of the pure system in thermal
equilibrium accomodates fluid, triangular solid
and square solid phases\cite{2dmod}. In the driven
system, as $F$ is increased, analogous phases appear
in approximately the inverse order to the sequence
obtained in the pure case as $T$ is increased.
This observation agrees roughly with the KV proposal
\cite{vinok}, which identifies
the shaking temperature $T_s$
with $T$.  
The shaking temperatures are predicted to fall
as $\sim 1/v^2$ and as $\sim 1/v$ in the drive and
the transverse directions respectively, consistent
with our observations in Fig.~\ref{shakord}(a).  We find
that $T^\alpha_s$ is nearly independent of $v_3$.
Importantly, within the putative coexistence regime,
$T^\alpha_s$ behaves non-monotonically, implying a breakdown
of the KV prediction (see Fig.~\ref{shakord}(a)).
Typically, for a particular disorder configuration
and for $5.5 < v_3 < 8.5$, $T^\alpha_s$ appears to increase
sharply at a well defined $F$, signifying the start of
coexistence. Within G, $T^\alpha_s$ remains high but drops
sharply at the upper limit of G, to {\em continue to
follow the interrupted KV behavior}. The limits of
the coexistence region, though sharp for any typical
disorder realization, vary considerably {\em between}
realizations.

Within the coexistence region the probability of
obtaining triangular (square) regions appears to
decrease (increase) roughly linearly with $F$;
see Fig.~\ref{shakord}(c).  Real
space configurations (Fig.~\ref{config}) exhibit islands
of square and triangular coordination connected
by interfacial regions with predominately 5
coordinated particles. Particles with coordination
7 are typically associated with dislocations which
are scattered randomly in the interface. This
configuration, in the co-moving frame, is extremely
dynamic, with the islands rapidly interconverting
between square and triangle.  This interconversion
has complex temporal properties: the power spectrum
of coordination number fluctuations shows a prominent
$1/f$ fall off over several decades. In addition,
particle curent fluctuations are {\em enhanced} by
an order of magnitude, also displaying a regime of
$1/f$ behavior (Fig.~\ref{autocor}), although over
a restricted range as a consequence of the proximity
to the washboard frequency. This result would be hard
to interpret in the absence of a genuine coexistence
phase: above the depinning transition, increasing the
driving force would naively be expected to {\em reduce}
current noise monotonically, as observed in all
previous simulation work on related models
\cite{faleski,vort-dyn,fangohr} 

Renormalization group arguments suggest that neither
translational LRO nor QLRO survive in the disordered
moving state in {\em two} dimensions at any finite
drive\cite{balents,scheidl,gia2}, their closest analog
being the moving Bragg glass state argued to be
stable in three dimensions and higher\cite{gia2}.
The ordered square (F) and triangular (E)
states we obtain at large drives are then to be
understood as a finite size effect arising from the
restricted size of our simulation box, although the
crossover length scales can be very large at weak
disorder\cite{balents,gia2}.  The possibility of
alternative dynamically stabilized states with reduced
levels of ordering, such as driven transverse smectics,
is attractive\cite{balents,nattermann}. In contrast
to some previous work\cite{moon,vort-dyn,fangohr}, we
see no evidence for smectic order and flow in weakly
coupled channels at large drives -- our channels always
remain strongly coupled --  but note that moving states
in which channels transverse to the drive direction
are effectively decoupled may be stabilized at higher
levels of thermal noise or  randomness\cite{fangohr}.

No general arguments seem to rule out QLRO in the
``{\em moving hexatic glass}'' phase in two dimensions and
our simulations support this possibility; see also
Ref.~\cite{dpt-hexatic}. The {\em
drive-induced stabilization of the triangular lattice
state} we obtain, even well into regimes where
$v_3$ would favour a square, is an unusual result.
Our finding of a distinct {\em coexistence regime}
(G) separate from plastic flow states, hexatic
and crystal, is novel. Such coexistence occurs in a
narrow and reproducible regime in parameter space and
is characterized by a variety of dynamic anomalies,
including enhanced noise signals with $1/f$ character.
The coexistence state appears to be a genuine
non-equilibrium state, separated from other regimes
through sharp non-equilibrium transitions. 

Theoretical work has, so far, neglected the
possibility of such dynamic phase coexistence
in the non-equilibrium steady states of driven
disordered crystals\cite{nattermann}. A non-disordered
but frustrated system closely related to the one
considered here has been proposed recently as a model
for the dynamical heterogeniety seen in the glassy
state\cite{nature}. In this model, fluctuating regions
of crystalline ordering within a liquid background are
argued to be responsible for the anomalous dynamical
behaviour and slow relaxation in the glassy state,
a physical picture which shares some similarities to
our ideas regarding the coexistence regime.

In conclusion, we have shown here that the competition
between structural phase transitions in a pure system
as modified by disorder, coupled to the non-equilibrium
effects of an external drive, can have a variety of
non-trivial consequences. The ubiquity of structural
phase transitions in the vortex state of a large number
of superconductors which have been studied recently,
as well as the relative ease with which the vortex
state can be driven, suggests experimental situations
in which the ideas here should find application.
Functionalized colloidal particles driven over random substrates
constitute a novel system on which our proposals can be
tested \cite{coll-rand,comp-coll}. Such systems have the further
advantage that interparticle interactions can be tuned
both through surface modifications and through the
application of external fields\cite{comp-coll}. 

The authors thank M. Rao, J. Bhattacharya 
and A. Chaudhuri for discussions. This work was partially 
supported by the DST (India).

%%%%%%%%%%%%%%%%%%%%%%%%%%%%%%%%%%%%%%%%%%%%%%%%%%%%%%%%%%%%%%%%%%%%%
%                     REFERENCE PAGE
%%%%%%%%%%%%%%%%%%%%%%%%%%%%%%%%%%%%%%%%%%%%%%%%%%%%%%%%%%%%%%%%%%%%%%


\begin{thebibliography}{99}

\bibitem{special} {\em Special Section on Nonequilibrium Statistical Systems},
M. Barma {\it ed.}, Current Science {\bf 77} (3), 402 (1999).

\bibitem{gia} T. Giamarchi and S. Bhattacharya, in {\em High Magnetic Fields: 
Applications in Condensed Matter Physics and Spectroscopy}, 
ed. C. Berthier {\it et al.}, Springer-Verlag, (2002), p 314.

\bibitem{super} C. D. Dewhurst, S. J. Levett, and D. McK. Paul, Phys. 
Rev. B {\bf 72}, 014542 (2005); L. Ya. Vinnikov {\it et al.},  
Phys. Rev. B {\bf 64}, 220508(R) (2001); B. Rosenstein {\it et al.}, 
Phys. Rev. B {\bf 72}, 144512 (2005); S.P. Brown {\it et al.}, Phys. 
Rev. Lett. {\bf 92}, 067004 (2004); R. Gilardi {\it et al.}, 
Phys. Rev. Lett. {\bf 93}, 217001 (2004)

\bibitem{anand} A. Yethiraj and A. van Blaaderen, Nature {\bf 421}, 513 (2003);
A. Yethiraj {\it et al.}, Phys. Rev. Lett. {\bf 92}, 058301 (2004)

\bibitem{coll-rand}A. Pertsinidis and X. S. Ling, Bull. Am. Phys. Soc. 
{\bf 46}, 181 (2001);{\em ibid}, {\bf 47}, 440 (2002); C. Reichhardt and 
C. J. Olson, \prl, {\bf 89}, 078301, (2002). 

\bibitem{2dmod} M. Rao and S. Sengupta, J. Phys: Condens. Mat. {\bf 16},
7733 (2004), \prl {\bf 91}, 045502 (2003);

\bibitem{Chud} E. M. Chudnovsky and R. Dickman, Phys. Rev. B, {\bf 57},
2724, (1998); A. Sengupta, S. Sengupta and G.I. Menon, 
Europhys. Lett., {\bf 70}, 635 (2005).

\bibitem{delaunay} F.R. Preparata and M. I. Shamos,
{\em Computational Geometry: An Introduction},
Springer-Verlag, New York (1985)

\bibitem{vinok}A. E. Koshelev and V. M. Vinokur, \prl {\bf 73}, 3580 (1994);
S. Scheidl and V. M. Vinokur, \prb {\bf 57}, 13800, (1998).

\bibitem{jensen} H. J. Jensen, A. Brass and A. J. Berlinsky,
Phys. Rev. Lett. {\bf 60}, 1676 (1988)

\bibitem{moon} K. Moon, R.T. Scalettar and G. T. Zimanyi, Phys.
Rev. Lett. {\bf 77}, 2778 (1996)

\bibitem{faleski} M.C. Faleski, M.C. Marchetti and A. A. Middleton,
Phys. Rev. B {\bf 54}, 12427 (1996)

\bibitem{vort-dyn} C. J. Olson, C. Reichhardt 
and F. Nori, \prl {\bf 81}, 3757, (1998).

\bibitem{fangohr}H. Fangohr, S. J. Cox and P. A. J. de Groot \prb, {\bf 64},
064505, (2001)

\bibitem{chandran} M. Chandran, R.T. Scalettar and 
G.T. Zimanyi, Phys. Rev. B {\bf 67}, 052507 (2003)

\bibitem{dpt-hexatic}S. Ryu, A. Kapitulnik, and S. Doniach, Phys. Rev. Lett.  
{\bf 77}, 2300 (1996).

\bibitem {largev3} Although the plastic flow regime expands 
as $v_3$ is increased over the scales shown in Fig.~1, it 
collapses at much larger $v_3 \simeq 30$ (not shown).

\bibitem{balents} L. Balents, M. C. Marchetti and L. Radzihovsky, 
Phys. Rev. Lett. {\bf 78}, 751 (1997); Phys. Rev. B {\bf 57}, 7705 (1998).

\bibitem{scheidl} S. Scheidl and V. M. Vinokur, \pre {\bf 57}, 2574, (1998).

\bibitem{gia2} P. LeDoussal and T. Giamarchi, \prb {\bf 57}, 11356 (1998).

\bibitem{nattermann} T. Nattermann and S. Scheidl, Adv. in Phys.
{\bf 49}, 607 (2000).

\bibitem{nature} H. Shintani and H. Tanaka, Nature Physics
{\bf 2} 200 (2006) 

\bibitem{comp-coll}A. van Blaaderen, Nature, {\bf 439}, 545, (2006).

\end{thebibliography}
\end{document}